\documentclass[12pt]{article}
\usepackage{amsmath,amssymb}
\usepackage[dvips]{graphicx}

\def\CC{{\cal C}}
\def\CD{{\cal D}}

\newcommand{\bbibitem}[1]{\bibitem{#1}\marginpar{#1}}

\def\Label#1{\label{#1}%
  \smash{\hbox to0pt{\raise1ex\hbox{\tiny[#1]}\hss}}}
\def\noLabels{\let\Label=\label}
\def\nobbibitem{\let\bbibitem=\bibitem}

\newcommand{\be}{\begin{equation}}
\newcommand{\ee}{\end{equation}}
\newcommand{\bea}{\begin{eqnarray}}
\newcommand{\eea}{\end{eqnarray}}

\newcommand{\nmax}{{n_{max}}}
\begin{document}

\rightline{HIP-2007-11/TH}
\vskip 1cm \centerline{\large {\bf Disk
Partition Function}} \centerline{ \large {\bf and Oscillatory Rolling
Tachyons}}
\vskip 1cm
\renewcommand{\thefootnote}{\fnsymbol{footnote}}
\centerline{{\bf Niko Jokela,$^{1}$\footnote{niko.jokela@helsinki.fi} Matti
J\"arvinen,$^{1,2}$\footnote{mjarvine@ifk.sdu.dk}}}
\centerline{{\bf Esko
Keski-Vakkuri,$^{1,2}$\footnote{esko.keski-vakkuri@helsinki.fi} and
Jaydeep Majumder$^{1}$\footnote{jaydeep.majumder@helsinki.fi} }}
\vskip .5cm \centerline{\it ${}^{1}$Helsinki Institute of Physics
and ${}^{2}$Department of Physical Sciences } \centerline{\it
P.O.Box 64, FIN-00014 University of Helsinki, Finland}

\setcounter{footnote}{0}
\renewcommand{\thefootnote}{\arabic{footnote}}

\begin{abstract}
An exact cubic open string field theory rolling tachyon solution was recently found by
Kiermaier et. al. and Schnabl. This oscillatory solution has been argued
to be related by a field redefinition to the simple exponential
rolling tachyon deformation of boundary conformal theory. In the latter
approach, the disk partition function takes a simple form. Out of curiosity,
we compute the disk partition function for an oscillatory tachyon profile,
and find that the result is nevertheless almost the same.
\end{abstract}

\newpage

\section{Introduction}

Recently there has been remarkable new
analytic progress in the study of cubic open string field theory (OSFT) \cite{Witten:1985cc}.
In particular, an exact rolling tachyon solution was found
\cite{Kiermaier:2007ba}, related to tachyon matter
and decay of an unstable D-brane. The profile of the tachyon component
of the full string field obtained by
\cite{Kiermaier:2007ba} from Witten's cubic OSFT is
\be\label{tachyon}
 T_\lambda(x^0) = \lambda e^{\frac{1}{\sqrt{\alpha'}} x^0}+\sum_{n=2}^{\infty}(-1)^{n+1}\lambda^n
 \beta_n e^{\frac{1}{\sqrt{\alpha'}}n x^0} \ ,
\ee where $\beta_n$ are positive coefficients\footnote{We follow the
convention where the true minimum of the tachyon effective potential
is at some $T>0$ while keeping $\lambda > 0$. We work in units where
$\alpha'=1$.} with a known integral representation. The authors of
\cite{Kiermaier:2007ba} started from the exactly marginal operator
\be \label{tachyoncft}
V = e^{\frac{1}{\sqrt{\alpha'}} X^0} \ ,
\ee
constructed the full OSFT solution recursively, adopting the gauge
choice of
\cite{Schnabl:2005gv}, and obtained (\ref{tachyon}). Generalizations to
superstrings have been reported in \cite{susywork},
and related work is also \cite{Fuchs:2007yy}.

The solution (\ref{tachyon}) has an oscillatory structure, as was suggested to be characteristic
for the OSFT rolling tachyon by the
previous investigations \cite{Coletti:2005zj,previouswork}. On the other hand, in the boundary
conformal field theory (BCFT) description of the same process\footnote{For another reference
on the relation between SFT solutions and deformations of BCFT, see \cite{Kluson:2003xu}.}, the tachyon field rolls monotonously,
represented by the simple exponential (\ref{tachyoncft}). The apparent contradiction was addressed
in \cite{Coletti:2005zj}. The OSFT string field solution
contains an infinite tower of other (massive) fields which are
sourced by the rolling tachyon component.
One can perform a field redefinition to boundary string field theory (BSFT)
\cite{bsft}\footnote{A pedagogical discussion of BSFT is also \cite{Kraus:2000nj}.} variables,
in such a way that all other fields except the tachyon are zero \cite{kmm}. In the BSFT field
coordinatization the tachyon can then turn out to be the simple exponential (\ref{tachyoncft}), while it
was oscillatory in the OSFT frame \cite{Coletti:2005zj}. Thus the marginal
OSFT solution (where the tachyon component is off-shell) maps to a manifestly on-shell form. Further, it
maps to the exactly marginal operator which gives a BCFT deformation. For the new full OSFT solution of
\cite{Kiermaier:2007ba} this was shown in \cite{Ellwood:2007xr}.
Since the new rolling tachyon solution relates to the known BCFT deformation, in particular the time evolution
of pressure of the associated
tachyon matter has already been calculated in \cite{SLNT}, it corresponds to the disk partition function
of the BCFT with $\lambda V$ (\ref{tachyoncft}),
\be\label{psimple}
 p(x^0) = Z_{{\rm disk}}(x^0) = \frac{1}{1+2\pi \lambda e^{x^0}} \ .
\ee

In this note, we are reporting a curious observation. Suppose we were to consider BSFT with an oscillatory
off-shell tachyon profile of the form (\ref{tachyon}). Consider the worldsheet CFT and turn on the boundary
the tachyon field (\ref{tachyon}),
\be\label{defo}
  S = S_0 + \oint_{\partial \Sigma} dt~T_\lambda (X^0(t)) \ ,
\ee
it is off-shell and breaks the conformal invariance on the boundary. Suppose we attempt to
do a straightforward calculation of the disk partition
function, leaving the zero mode $x^0$ unintegrated. Given the oscillatory behaviour of (\ref{tachyon}),
we would probably expect the resulting disk partition function to be quite unwieldy and very different
{}from (\ref{psimple}).

However, when we perform the
string worldsheet theory analysis (along the lines
of \cite{SLNT}),
surprisingly we find that the result is {\em almost the same} as (\ref{psimple}), with maximum 1\% relative
deviation. The deviation only appears at times close to the value $x^0 \sim -\ln 2\pi \lambda$. Apart
from the deviation, there is no oscillatory behaviour -- at late times the disk partition functions become
identical.
We do not
quite know how to interpret this curious observation. Apparently the field redefinitions involved in
mapping from the oscillatory tachyon profile to the monotonously rolling one are not always so significant
from the point of view of interesting observables. Further, while in our calculation the tachyon is
of the form (\ref{tachyon}), the actual values of the coefficients $\beta_n$ do not matter much --
in particular they (and the tachyon field) need not be the same as in \cite{Kiermaier:2007ba}.
Interpretational issues aside, we believe that
the calculational tricks which we have used will be useful for other investigations and thus interesting
in their own right.

\section{The disk partition function}

In the first quantized string worldsheet approach, we turn on the tachyon background
(\ref{defo}).
The disk
partition function is (separating out the zero mode $X^0 = x^0 +
X'^0$ and leaving it unintegrated)
\be\label{Zopen}
 Z_{\rm{disk}}(x^0) 
 =\int \CD X'^0 \CD\vec X e^{-S_0}\exp \left( -\oint_{\partial \Sigma} dt~T_\lambda (x^0+X'^0(t)) \right) \ .
\ee
Note that, in the limit $\beta_{n>1}\to 0$, we expect to produce the familiar results for
half S-brane \cite{SLNT}.

By expanding in the boundary perturbation in (\ref{Zopen}) as a power series,
and carefully following the calculational steps outlined in
\cite{Balasubramanian:2004fz},
the disk partition function is
\bea\label{pf}
 Z_{\rm{disk}}(x^0) 
 & = & \prod_{n=1}^{\infty}\sum_{N_n=0}^{\infty}
 \frac{((-1)^n\lambda^n \beta_n e^{nx^0})^{N_n}}{N_n!}
\int dt^{(n)}_1\cdots dt^{(n)}_{N_n} \langle \prod_{n,i} e^{n
X'^0(t^{(n)}_i)}\rangle\nonumber\\
& = & \sum_{\{N_1,N_2,\ldots\}=0}^\infty \left(\prod_{n=1}^\infty\frac{((-1)^n z_n)^{N_n}}{N_n!}\right)
\cdot I(N_1,N_2,\ldots) \ ,\label{pdef}
\eea
with
\be
z_n \equiv 2\pi\lambda^n\beta_n e^{n x^0} > 0
\ee
and $\beta_1=1$, and where
\bea \label{Idef}
 I(N_1,N_2,\ldots) &\equiv & \int\left[\prod_{n=1}^\infty\prod_{i=1}^{N_n}\frac{dt_i^{(n)}}{2\pi}\right]
 \left[\prod_{n=1}^\infty\prod_{1\leq i<j\leq N_n}|e^{it_i^{(n)}}-e^{it_j^{(n)}}|^{2n^2}\right] \\
 & & \cdot\left[\prod_{1\leq n<m}^\infty\prod_{i=1}^{N_n}\prod_{j=1}^{N_m}
 |e^{it_i^{(n)}}-e^{it_j^{(m)}}|^{2nm}\right] \label{eq:integrals} \nonumber
\eea
denotes an infinite product of coupled integrals.

The above formulas are just formal expressions, before
good domains of convergence are found.
It is difficult to analyze the problem fully -- so we will
first study a simpler toy model.

\section{A warm-up calculation: the Dyson series}

We have two tasks at hand: (i) to try to calculate the integrals
(\ref{Idef}) and (ii) to try to control the series (\ref{pf}). These
tasks appear to be rather challenging, so we will first consider a
toy model calculation. It is reminiscent of the actual one but
allows us to carry out both tasks.

We consider a series expansion, which we will call the ``Dyson
series'' from now on. It is inspired by the integration formula to
compute the canonical partition function of a Dyson gas
\cite{Dyson:1962es}, \be\label{dyson}
 \int \prod_{i=1}^N \frac{dt_i}{2\pi} \left[ \prod_{i<j} |e^{it_i} - e^{it_j}|^\beta \right]
 = \frac{\Gamma(1+\frac{\beta N}{2})}{[\Gamma(1+\frac{\beta}{2})]^{N}} \ ,
\ee for which various proofs have been presented in the literature (see
\cite{mehta}). The integral (\ref{Idef}) resembles an infinite
product of decoupled Dyson gas integrals (\ref{dyson}), except for the last
cross coupling term in the square brackets in the integrand of (\ref{Idef}). Let
us first truncate the infinite product and keep just $n_{max}$
first terms, with integer $n_{max}\gg1$. (In the end we will
consider the limit $n_{max}\rightarrow \infty$.) Then, consider
the cross coupling term in the integrand of
(\ref{Idef}), which renders the
integral difficult to evaluate. Let us rewrite it as
\be
\prod_{1\leq n<m}^\nmax\prod_{i=1}^{N_n}\prod_{j=1}^{N_m}
 |e^{it_i^{(n)}}-e^{it_j^{(m)}}|^{2nm}
 = \prod_{1\leq n<m}^\nmax\prod_{i=1}^{N_n}\prod_{j=1}^{N_m}
 \left(1-\frac{e^{it_j^{(m)}}}{e^{it_i^{(n)}}}\right)^{nm}
 \left(1-\frac{e^{it_i^{(n)}}}{e^{it_j^{(m)}}}\right)^{nm} \ .
\ee Now it turns out that the integral simplifies drastically if we
replace the exponent $nm$ in the first term on r.h.s. by $n^2$, and
the second exponent $nm$ by $m^2$. This step is clearly {\em ad
hoc}. However, it is a useful trick to try, since it simplifies the
calculations enough to give a tractable toy model calculation to
practice with and to gain insight for the actual disk partition
function calculation.
So we consider a version of the series (\ref{pf}) where we replace
the original integrals (\ref{Idef}) by \bea \label{Iapprox}
 \tilde I(N_1,N_2,N_3,\ldots;n_{max})
   & = & \int\left[\prod_{n=1}^{n_{max}}\prod_{i=1}^{N_n}\frac{dt_i^{(n)}}{2\pi}\right]
   \left[\prod_{n=1}^{n_{max}}\prod_{1\leq i<j\leq N_n}|e^{it_i^{(n)}}-e^{it_j^{(n)}}|^{2n^2}\right]
   \nonumber \\
   & & \cdot\left[\prod_{1\leq n<m}^{n_{max}}\prod_{i=1}^{N_n}\prod_{j=1}^{N_m}
   \left(1-\frac{e^{it_j^{(m)}}}{e^{it_i^{(n)}}}\right)^{n^2}
   \left(1-\frac{e^{it_i^{(n)}}}{e^{it_j^{(m)}}}\right)^{m^2}  \right]   \nonumber   \\
   & = & \frac{\Gamma (1+\sum_{n=1}^{n_{max}} n^2 N_n)}{\prod_{n=1}^{n_{max}}
    [\Gamma (1+n^2)]^{N_n}} \ ,
\eea where the last line is the exact analytical result for the
integral \cite{mehta,goodwilson}. Since the integrals
(\ref{Iapprox}) are a variation of the Dyson gas integral formula
(\ref{dyson}), we call the new series ``Dyson series''. In
Appendix A we compare the original integrals $I$ with the approximate
ones $\tilde I$, for some cases where it is possible to calculate
the original integral analytically, to see how much Dyson series toy
model deviates from (\ref{Iapprox}) the exact formula.

The virtue of the Dyson series is that we can also solve the task (ii): we
can actually sum the series in a controlled way. We will first
recognize it as an asymptotic series, but can rewrite it as an
integral formula which we can regulate by a suitable deformation of
integration contour. We will discuss that next.

\section{Summing the Dyson series}

Instead of the series (\ref{pf}) we consider the Dyson series with coefficients
$\tilde I$ instead of $I$.
We also simplified it further by truncating the infinite product, so that we
have
\bea \label{stsum}
 Z_{\rm Dyson}(x^0;n_{max}) &=& \left(\prod_{n=1}^{n_{max}}
 \sum_{N_n=0}^\infty\frac{((-1)^n z_n)^{N_n}}{N_n!}\right)
 \frac{\Gamma (1+\sum_{n=1}^{n_{max}} n^2 N_n)}{\prod_{n=1}^{n_{max}}
    [\Gamma (1+n^2)]^{N_n}} \ .
\eea
Even after truncating to a finite product of $\nmax$ terms, the expression is not well
behaved since the product is that of possibly divergent infinite series. In order to gain
better control, we rewrite (\ref{stsum}) as an integral representation,
\bea \label{integ}
Z_{\rm Dyson}(x^0;n_{max}) &=& \left(\prod_{n=1}^{n_{max}}\sum_{N_n=0}^\infty
 \frac{((-1)^n z_n)^{N_n}}{N_n!(n^2)!^{N_n}}\right)\int_0^\infty du
 u^{\sum_{n=1}^\nmax n^2 N_n} e^{-u} \nonumber\\
 &=&\int_0^\infty du\ \exp\left[-u + \sum_{n=1}^\nmax \frac{(-1)^n z_n u^{n^2}}{(n^2)!}\right] \ .
\eea
Now we have a single integral, and the exponent in the integrand is a finite
sum of $\nmax$ terms.  Let us take a closer look at it. We denote
\be
 F_\nmax(u) = -u + \sum_{n=1}^\nmax \frac{(-1)^n z_n u^{n^2}}{(n^2)!} \ .
\ee For real $u$, $F_\nmax (u)$ is oscillatory with the amplitude of
oscillation increasing with $u$. The largest oscillations are due to
the terms with $n \simeq \nmax$. As a consequence, the integral
(\ref{integ}) does not have the expansion (\ref{stsum}) for small
$z_n$, and the limit $\nmax \to \infty$ does not exist.
We will next give a prescription to regulate the integral.

Let us deform the contour of integration in (\ref{integ}) away from
the positive real axis. If the integrand would be analytic, this
would have no effect. However, it has an essential singularity at
infinity. Consequently, the contour deformation will change the
integral, due to a different approach to the point at infinity. Thus
we can regulate the integral (\ref{integ}) by finding a suitable
contour deformation. However, the integral will then also become
complex valued. Since the pressure is real valued, we adopt a
prescription where we define it to be the real part of the integral
over the deformed contour\footnote{With this prescription, it
reproduces the asymptotic series (\ref{stsum}). If one has a strong
preference to keep the integral real valued, one can alternatively
first write it as a sum of two identical terms, then deform the
contour in two opposite ways as mirror images of each other so that
the two terms become complex conjugates.}, \be\label{focus}
 Z_{\rm Dyson}(x^0;\nmax) =\mathrm{Re}  \int_{\CC} du\
\exp \left[ F_\nmax(u) \right] \ ,
\ee
where $\CC$ runs from $0$ to $\infty$ such that $\mathrm{Re} F_\nmax$ decreases monotonically on it.
For the choice of $\CC$, see Fig.~1 which depicts the eye-appealing structure of the real part
of $F_\nmax$ (the plot is shown for the value $\nmax=11$).
The regular structure of $\mathrm{Re}F_\nmax$ arises from the fact
that $\mathrm{Re}F_\nmax(u)$ is dominated by the $n$th term of the
sum at $|u| \simeq n^2$. Fig.~1 suggests that
there is a preferred choice for a path (in the quadrant $0<\phi<\pi/2$) from $0$
to $\infty$ that avoids all the light gray regions and proceeds in the direction of darker
color (decreasing $\mathrm{Re}F_\nmax$). We call such a path $\CC_{pref}$ and focus on
(\ref{focus}) with $\CC=\CC_{pref}$
which stays well defined in the limit $\nmax \to \infty$.

\begin{figure}[ht]
\begin{center}
\noindent
\includegraphics[width=0.6\textwidth]{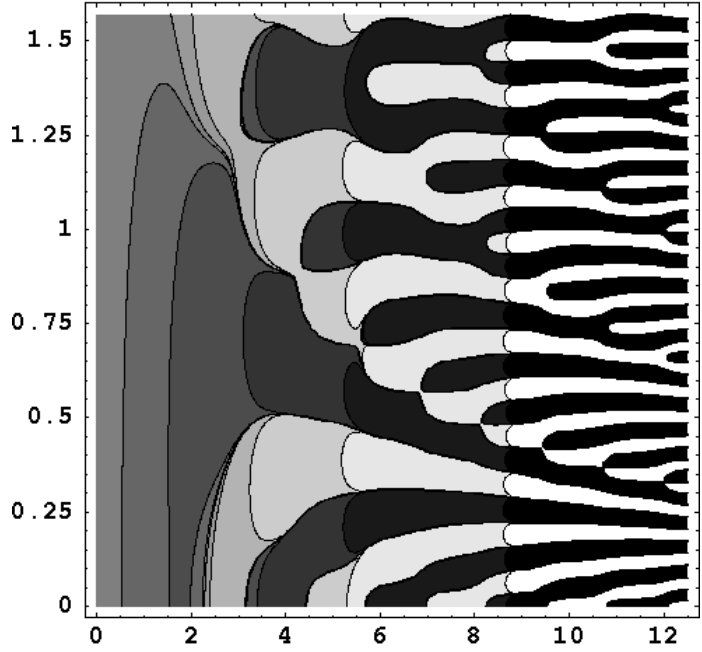}
\end{center}
\caption{$\mathrm{Re}F_{11}(r^2e^{i\phi})$ for $r=0 \ldots 12.5$ (horizontal axis) and for $\phi=0 \ldots \pi/2$ (vertical axis) with $x^0=0$. $\mathrm{Re}F_{11}$ is small in the dark regions.}
\label{fig:convergence}
\end{figure}

As an example, let us consider the leading correction with $\nmax=2$.
We take $\CC_{pref}$
with a constant phase, {\it i.e.}, $u=re^{i\pi/4}$ with $r=0 \ldots \infty$. Then
\be \label{corrn2}
  Z_{\rm Dyson}(x^0;\nmax=2) =\mathrm{Re}\int_0^\infty dr \exp\left[i\pi/4-(1+z_1)re^{i\pi/4} - z_2 r^4/24 \right] \ ,
\ee
which is well defined. (Recall that $z_n = z_n(x^0) \sim \exp (nx^0)$.)
Developing the integrand at $z_2=0$ we get back the (asymptotic) series
\be
 Z_{\rm Dyson}(x^0;2) = \frac{1}{1+z_1}+\frac{z_2}{(1+z_1)^5} + \frac{35 z_2^2}{(1+z_1)^9} + \cdots \ .
\ee
We want to compare this to the leading term
\be\label{Z1disk}
 Z_{\rm Dyson}(x^0;1)\equiv \frac{1}{1+z_1} = \frac{1}{1 + 2\pi \lambda e^{x^0}} \ .
\ee
Numerical integration of (\ref{corrn2}) verifies that the {\em total}
correction with $\nmax=2$ is small:
\be
 \left[ \frac{Z_{\rm Dyson}(x^0;2)-Z_{\rm Dyson}(x^0;1)}{Z_{\rm Dyson}(x^0;1)}\right]_{max, x^0 \in R}
 \sim 10^{-3}
\ee
and well described by the first few terms of the asymptotic series.
Note then that at late times the first subleading term is $\sim z_2z^{-5}_1 \sim e^{-3x^0}$, which
is much smaller than the leading term $\sim z^{-1}_1 \sim e^{-x^0}$.
One can argue that at late times $x^0 \to \infty$ all subleading terms
are negligible compared to the leading $e^{-x^0}$ behavior.
Similarly one finds that the $\nmax=3$ correction is even smaller
\be
 \left[ \frac{Z_{\rm Dyson}(x^0;3)-Z_{\rm Dyson}(x^0;2)}{Z_{\rm Dyson}(x^0;1)}\right]_{max, x^0 \in R} \sim 10^{-7} \ .
\ee
Refining the approximation to larger values of  $\nmax$ produces even more negligible corrections.
Thus the total correction to the leading result (\ref{Z1disk})
is at most $\sim 10^{-3}$ in the Dyson series, even when $\nmax \to \infty$. Thus, the
approximate result for the disk partition function decays exponentially at late times,
\be
  Z_{\rm Dyson}(x^0) = \lim_{\nmax \to \infty} Z_{\rm Dyson}(x^0,\nmax)~\sim~e^{-x^0} \
  \ ; \ x^0 \rightarrow \infty \ .
\ee

We will now return back to our original problem, the disk partition function (\ref{pf}). The lesson from the
Dyson series toy model is that it is useful to truncate the infinite products by introducing a `cut-off'
$n_{max}$ and then try to see how much the time dependence is corrected as $n_{max}$ is increased. If
the additional corrections are more and more subleading, they can be ignored in the limit $n_{max}\rightarrow
\infty$. The full series is in fact well approximated by just the leading terms as $x^0\rightarrow \infty$.
The partition function (\ref{pf}) turns out to have a similar behavior.

\section{The original disk partition function at late times}

Consider again the exact series (\ref{pf}).
In our toy model the relevant late-time corrections are produced by the first terms
in the asymptotic series (\ref{stsum}).
It turns out that the first terms of the exact series (\ref{pf}) can also be calculated analytically,
without using any approximation for $I$.  The first correction terms
are those, where most of the $N_2,N_3,\ldots$ are zero. We denote the integral coefficients
of these by
\be
 I_n(N_1,N_n) \equiv I(N_1,0,0,\ldots,0,N_n,0,0,\ldots)
\ee
so, {\em e.g.}, $I_2(N,4)=I(N_1=N,N_2=4,0,0,\ldots)$. It turns out we can evaluate the integrals
\be
 I_n(N,1) = \int \frac{dt_1^{(n)}}{2\pi} \prod_{i=1}^N \frac{dt_i^{(1)}}{2 \pi}\prod_{i<j} \left|e^{it_i^{(1)}}-e^{it_j^{(1)}}\right|^2 \prod_i\left|e^{it_i^{(1)}}-e^{it_1^{(n)}}\right|^{2n} \ .
\ee
This is a well-known Selberg integral, and has previously been applied
in the context of rolling tachyons in \cite{Balasubramanian:2004fz}. The result reads
\be \label{Selres}
 I_n(N,1) = N! \prod_{j=1}^N \frac{\Gamma(j)\Gamma(j+2n)}{\Gamma(j+n)^2} = N!\prod_{j=0}^{n-1} \frac{j!}{(n+j)!}\frac{(N+n+j)!}{(N+j)!} \ .
\ee
In particular we find
\bea \label{Iident}
 \frac{I_2(N,1)}{N!} &=& \frac{N+2}{12}\frac{(N+3)!}{N!} = {N+4 \choose 4} + {N+3 \choose 4} \nonumber \\
&=&  \frac{1}{4!}\left[\frac{(N+4)!}{N!}+\frac{(N+3)!}{(N-1)!} \right] \ , \\ \nonumber
 \frac{I_3(N,1)}{N!} &=& \frac{1}{9!}\left[\frac{(N+9)!}{N!}+10 \frac{(N+8)!}{(N-1)!}+20\frac{(N+7)!}{(N-2)!}+10\frac{(N+6)!}{(N-3)!}+\frac{(N+5)!}{(N-4)!} \right] \ ,
\eea
where the first terms of the sums are the same as in the Dyson series toy model.

Thus, we find the corrections to $Z_{\rm disk}$ (eqn. (\ref{pf})) that are linear in $z_{2,3}$:
\bea
 Z_{\rm disk}(x^0) &=& \sum_{N=0}^\infty(-1)^N z_1^N\left[1 +  z_2 \frac{I_2(N,1)}{N!} - z_3\frac{I_3(N,1)}{N!} + \cdots \right] \nonumber \\
        &=& \frac{1}{1+z_1}+\frac{z_2(1-z_1)}{(1+z_1)^5} - \frac{z_3(1-10z_1+20z_1^2-10z_1^3+z_1^4)}{(1+z_1)^{10}} + \cdots \ .\label{lincorr}
\eea
From (\ref{Selres}) it follows that all higher order linear corrections (those depending on
$z_n$ with $n\geq 4$) have similar structures.

Note that the size of the corrections is slightly larger as in the Dyson series. In the latter,
at late times the correction linear in $z_2$ was $\sim z_2z^{-5}_1 \sim e^{-3x^0}$ but now
we find $\sim z_2z^{-4} \sim e^{-2x^0}$.  The correction linear in $z_3$ is subleading, we find
at late times $\sim z_3z^{-6}_1\sim e^{-3x^0}$.

Moving to higher order, the coefficients $I_2(N,2)$ apparently
also have a formula similar to (\ref{Iident}). We find
\be\label{eq:III}
 \frac{I_2(N,2)}{2!N!} = \frac{1}{8!}\left[35\frac{(N+8)!}{N!}+77\frac{(N+7)!}{(N-1)!}+27\frac{(N+6)!}{(N-2)!}+\frac{(N+5)!}{(N-3)!}\right] \ ,
\ee
whence the correction to the disk partition function that is quadratic in $z_2$ becomes
\be\label{nonlin}
 \frac{z_2^2(35-77z_1+27z_1^2-z_1^3)}{(1+z_1)^9} \ .
\ee
Interestingly, at late times this is of the same order as the linear correction,
namely $\sim z^2_2z^{-6}_1 \sim e^{-2x^0}$. As we will discuss below, at the order $z^n_2$ we will similarly
find $\sim z^n_2z^{-2n-2}_1\sim e^{-2x^0}$, and generalizing to order $z^n_3$ we will
find $\sim z^n_3z^{-3n-3}_1\sim e^{-3x^0}$. All these are small corrections compared to the
leading $\sim e^{-x^0}$ decay.

The above are still a tiny subset of all possible terms in the series (\ref{pf}), containing
all possible combinations of monomials of $z_1,z_2,z_3,\ldots$. But we can estimate
their late time behavior too.

Equations (\ref{Iident}), (\ref{eq:III}) show that the integers $I_n(N,1)$ and $I_2(N,2)$ can be expressed as finite sums over
binomial coefficients. Using methods outlined in appendix B, we evaluated 
\be
 \hat I(N_1,N_2,\ldots) =  \frac{1}{\prod_n N_n!}\ I(N_1,N_2,\ldots)
\ee
for almost all fixed values of $N_n$ for which $\hat I\lesssim 10^{19}$. 
Using these results we then discovered a generalizaton of the formulae (\ref{Iident}), (\ref{eq:III})
for more complicated
sets of $N_2, N_3, \ldots$. We find that
for any $N_1=N$ with fixed $N_2,N_3,\ldots,N_\nmax$
(with $N_\nmax>0$ and $0=N_{\nmax+1}=N_{\nmax+2}=\cdots$), the $\hat I$ can be written as a finite sum
\bea \label{genbin}
 \hat I(N_1=N,N_2,N_3,\ldots,N_\nmax) &=& \frac{1}{S!} \sum_{\ell=0}^{\ell_{max}} C_\ell~\frac{(N+S-\ell )!}
 {(N-\ell)!} \nonumber \\
 \mbox{} &=&
 \sum_{\ell=0}^{\ell_{max}} C_\ell {N+S-\ell \choose S}
\eea
where $S=\sum_{n=2}^\nmax n^2N_n$.
The relevant fact  for the moment is that the coefficients $C_\ell$
turn out to be independent\footnote{The formula (\ref{genbin})
has been evaluated and verified explicitly  (with explicit coefficients $C_\ell$),
\emph{e.g.}, for $(N_2,N_3,N_4)=(1,1,0)$, $(2,1,0)$, $(0,2,0)$ and $(1,0,1)$
in addition to the cases discussed above.}
of $N$. We will give an explicit formula for
$\ell_{max}$ below. The corresponding correction term
to $Z_{\rm disk}$ then becomes
\bea \label{gencorr}
\delta Z_{\rm disk} &=& \prod_{n=2}^\nmax \left[(-1)^n z_n\right]^{N_n} \sum_{N=0}^\infty (-z_1)^N
\hat I(N,N_2,N_3,\ldots,N_\nmax) \nonumber \\
&=& \prod_{n=2}^\nmax \left[(-1)^n z_n\right]^{N_n} \sum_{\ell=0}^{\ell_{max}} C_\ell
      \sum_{N=0}^\infty {N+S-\ell \choose S} (-z_1)^N \nonumber \\
&=& \prod_{n=2}^\nmax \left[(-1)^n z_n\right]^{N_n} \sum_{\ell=0}^{\ell_{max}}C_\ell \frac{(-z_1)^\ell}
{(1+z_1)^{S+1}} \ .
\eea

Importantly, for $\ell_{max}$ we found\footnote{Using (\ref{Selres}) it is straightforward to determine $\ell_{max}$ for the corrections which are linear in $z_n$ (with arbitrary $n=\nmax$). The general formula (\ref{ellmax}) was found by first making an educated guess
and then testing it with computer calculations. So far we have explicitly verified it up to $n_{max} = 4$ but
have not yet been able to construct a general proof.}
an explicit formula
\be\label{ellmax}
\ell_{max} =\sum_{n=2}^\nmax\left[n(n-1)N_n\right] -\nmax +1 \ .
\ee
The combination of (\ref{ellmax}) and the schematic formula (\ref{gencorr}) allows us to estimate the leading
late time dependence of all the correction terms to $Z_{\rm disk}(x^0)$. At late times the leading part
of the generic monomial correction (\ref{gencorr}) is given by the term with the highest exponent of
$z_1$, {\em i.e.}, the $\ell = \ell_{max}$ term. Then, combining the late time dependences
\bea
  \prod_{n=2}^{n_{max}} z_n^{N_n} &\sim & \exp [(\sum_{n=2}^{n_{max}} nN_n)x^0] \nonumber \\
  z_1^{\ell_{max}} &\sim & \exp [(\sum_{n=2}^{n_{max}} n(n-1)N_n)x^0 -(n_{max}-1)x^0] \nonumber \\
  z_1^{-(S+1)} &\sim & \exp [-(\sum_{n=2}^{n_{max}}n^2N_n)-x^0] \ ,
\eea
we find that the correction term (\ref{gencorr}) behaves as
\be
 \delta Z_{\rm disk} \sim  e^{-\nmax x^0}
\ee
at late times $x^0 \to +\infty$ and is thus subleading. Thus the leading correction is at most of
the order $e^{-2x^0}$.

\section{Summary}

\begin{figure}[ht]
\begin{center}
\noindent
\includegraphics[width=0.5\textwidth]{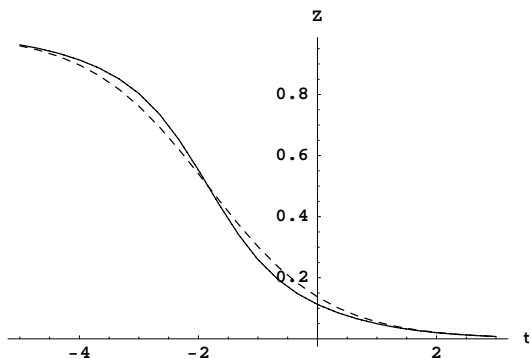}
\end{center}
\caption{The disk partition function (\ref{pcorr}) as a function
of time $t$. Here $\lambda=1$ and we used a large value $\sim 15$ for $\beta_2$.
For reference, the dashed line represents (\ref{psimple}).}
\label{fig:dpf}
\end{figure}

We have calculated the disk partition function with an oscillatory
tachyon field profile (\ref{tachyon}) instead of the exactly
marginal deformation (\ref{tachyoncft}). The largest
deviations, that we have found, from the disk partition function
(\ref{psimple}) of the latter
are surprisingly small, given by (\ref{lincorr}) and
(\ref{nonlin}). Including the largest one (linear in $z_2$) the
disk partition function reads
\bea \label{pcorr}
 Z_{\rm disk}(x^0) &\simeq& \frac{1}{1+2\pi\lambda e^{x^0}} + \frac{z_2(1-z_1)}{(1+z_1)^5} \nonumber\\
       &=&\frac{1}{1+e^{\tilde x^0}}
       + \frac{\beta_2}{2\pi}\frac{(e^{2\tilde x^0}-e^{3\tilde x^0})}{(1+e^{\tilde x^0})^{5}} \ ,
\eea where $\tilde x^0=x^0+\ln 2\pi \lambda$. Fig. \ref{fig:dpf}
shows the disk partition function with $\lambda=1$ and with
a large value of $\beta_2 \simeq 15$
for better visualization.
All the deviations seem to contribute around $x^0=-\ln 2\pi
\lambda$ and become smaller in size.
We find the result surprising: the disk partition function is very
similar to (\ref{psimple}) although the tachyon profile
(\ref{tachyon}) is oscillatory and very different from the monotonously
rolling (\ref{tachyoncft}). In particular, the oscillatory
behavior is almost washed out.

\newpage

\bigskip
\bigskip

\noindent
{\bf \large Acknowledgments}

\bigskip

We thank Asad Naqvi for useful discussions and Barton Zwiebach for
critical remarks on the earlier draft version of this work.  N.J. and M.J. have been in part supported by
the Magnus Ehrnrooth foundation. M.J. also acknowledges a grant from GRASPANP,
the Finnish Graduate School in Particle and Nuclear Physics.
This work was also partially supported by the EU 6th Framework Marie Curie Research and Training
network ``UniverseNet'' (MRTN-CT-2006-035863).

\bigskip

\bigskip

\noindent
{\large \bf APPENDIX A: A simple method for evaluating $I$}

\bigskip

Let us study an integral of the form
\be \label{Jndef}
 J_m = \int \prod_{i=1}^m \frac{dt_i}{2 \pi} \prod_{1\leq i<j\leq m} \left|e^{it_i}-e^{it_j}\right|^{2 k_{ij}} \ ,
\ee
where $k_{ij}$ are integers. This form is a generalization of (\ref{Idef}), where the exponents $n^2$ and $nm$ are allowed to take any values. The integral may be expressed as a finite sum by doing a Fourier transform. The ``propagator'' from $t_i$ to $t_j$ reads
\be
 S(t_j-t_i) = \left|1-e^{i(t_j-t_i)}\right|^{2 k_{ij}} = \sum_{n_{ij}=-k_{ij}}^{k_{ij}}(-1)^{n_{ij}}{2 k_{ij} \choose k_{ij}+n_{ij} } e^{in_{ij}(t_j-t_i)} \ .
\ee
By inserting this to (\ref{Jndef}) and by doing the $t$ integrals we have
\be
J_m = \left[\prod^m_{i < j}\sum_{n_{ij}=-k_{ij}}^{k_{ij}}(-1)^{n_{ij}}{2 k_{ij} \choose k_{ij}+n_{ij} }\right]\prod_{i=1}^m
\delta\left(\sum_{j=1}^{i-1}n_{ji}=\sum_{j=i+1}^{m}n_{ji}\right) \ ,
\ee
where only $m-1$  of the conditions in the (Kronecker)  delta functions are independent.
They can be used to fix the values of $n_{12},n_{13},\ldots$ so that
\be
J_m = \left[\prod_{1< i < j \le m}\sum_{n_{ij}=-k_{ij}}^{k_{ij}}(-1)^{n_{ij}}{2 k_{ij} \choose k_{ij}+n_{ij} }\right]\prod_{j=2}^m{2 k_{1j} \choose k_{1j}-\sum_{i=2}^{j-1} n_{ij}+\sum_{i=j+1}^m n_{ji}} \ .
\ee
This formula can be used to evaluate $I$ for small $n$ and $N_n$. {\it E.g.}, $I_2(2,2)$ is found
by letting $m=4$, $k_{12}=1$, $k_{13}=k_{14}=k_{23}=k_{24}=2$, $k_{34}=4$.
Some values are tabulated in Table \ref{table:uno}. Note that
\be
 \tilde I_2(N_1,N_2) = \frac{(N_1+4N_2)!}{4!^{N_2}} \leq I(N_1,N_2) \ .
\ee 
In Appendix B we present a more efficient method of evaluating $I$.
\\

\begin{table}\caption{Comparison of $I_2$ and $\tilde I_2$.}\label{table:uno}
\begin{tabular}{|c|c||c|c|c|}
 \hline
  $N_1$ & $N_2$ & $I_2(N_1,N_2)$ & $\tilde I_2(N_1,N_2)$ & point of interest \\
\hline\hline
  0 & 0 & 1 & 1 & \\
  k & 0 & $k!$ & $k!$ & \\
  0 & k & $\frac{(4k)!}{4!^k}$ & $\frac{(4k)!}{4!^k}$ & \\
  1 & 1 & $\frac{4!}{2!^2}=6$ & 5 & \\
  2 & 1 & $\frac{5!}{3}=\frac{5!3!}{2!}=40$ & 30 & \\
  3 & 1 & $5^3\cdot 3\cdot 2^2=\frac{5\cdot 5!}{2}=\frac{5\cdot 6!}{3\cdot 2^2}=300$ & 210 & \\
  4 & 1 & $2^3 3^2\cdot 5\cdot 7 = 7!=2520$ & 1680 & \\
  1 & 2 & $7^2 2^4=\frac{7!^2 2!^4}{5!^2 3!^2}=784$ & 630 & $I(1,2)=[{9\choose 4}+\frac{14}{3}]I(1,1)$\\
  2 & 2 & $5\cdot 3^3\cdot 17\cdot 2^2=9180$ & 6300 & \\
  3 & 2 & $2^4\cdot 3\cdot 2371=113808$ & 69300 & \\
  1 & 3 & $3\cdot 2^5 7^2 11^2=569184$ & 450450 & $I(1,3)=[{13\choose 4}+11]I(1,2)$\\
  2 & 3 & $2^6 3^3 7\cdot 13\cdot 61 = 9592128$ & 6306300 & \\
  1 & 4 & $2^{10}3^2 5^2 7^2 11^2 = 1366041600$ & 1072071000 & $I(1,4)=[{17\choose 4}+2^2 5]I(1,3)$\\
\hline
\end{tabular}
\end{table}

\bigskip

\noindent
{\large \bf APPENDIX B: A formula for $I$ using matrix determinants}

\bigskip

In this appendix the integral
\bea\label{Intdef}
 I(N_1,N_2,N_3\ldots) &=& \int\left[\prod_{n=1}^\infty\prod_{i=1}^{N_n}\frac{dt_i^{(n)}}
 {2\pi}\right]
 \left[\prod_{n=1}^\infty\prod_{1\leq i<j\leq N_n}|e^{it_i^{(n)}}-e^{it_j^{(n)}}|^{2n^2}\right] \nonumber \\
 & & \cdot\left[\prod_{1\leq n<m}^\infty\prod_{i=1}^{N_n}\prod_{j=1}^{N_m}|e^{it_i^{(n)}}-e^{it_j^{(m)}}|^{2nm}
 \right] 
\eea
is transformed to a finite sum over certain integer valued functions. This sum can then be used to evaluate $I$ exactly for a given set of $\{N_n\}$.

For $n=1$ (i.e., $0=N_2=N_3=\cdots$) (\ref{Intdef}) becomes
\be
 I_N = \int\prod_i \frac{dt_i}{2\pi}\prod_{1\leq i<j\leq N}|e^{it_i}-e^{it_j}|^2 \ .
\ee
Here the integrand is the absolute value squared of the Vandermonde determinant
\bea \label{Ddef}
 |\Delta(z_1,\ldots,z_N)|^2 &=& \prod_{1\leq i<j\leq N}|e^{it_i}-e^{it_j}|^2 =\left|\sum_{\{i\}}\varepsilon_{i_1 \cdots i_N} z_1^{i_1-1} \cdots z_N^{i_N-1}\right|^2 \nonumber\\
 &=& \left|\sum_\Pi (-1)^{\Pi} \prod_{k=1}^N z_k^{\Pi(k)-1}\right|^2
\eea
where $z_k=\exp(it_k)$ and $\Pi$ denotes permutations of ${1,2,\ldots,N}$.
It is easy to check that if (\ref{Ddef}) is expressed as a polynomial of $\{z_k\}$,
the constant term in the polynomial is equal to $I_N$.

The Vandermonde approach can be generalized for $n>1$ using confluent Vandermonde matrices.
This can be done by differentiation.
For example, 
\bea
 \prod_{1\leq i<j\leq N}|z_i-z_j|^{2n_in_j} &=& \left|\frac{\partial}{\partial z_{N+1}} \Delta(z_1,\ldots,z_{N},z_{N+1})\big|_{z_{N+1}=z_{N}}\right|^2 \\ \nonumber
&=& \left|\sum_{\{i\}} \varepsilon_{i_1 \cdots i_{N+1}} z_1^{i_1-1} \cdots z_N^{i_N-1}(i_{N+1}-1)z_N^{i_{N+1}-2}\right|^2
\eea
where $n_{I_N}=2$ and all other $n_i=1$. This is the determinant of a confluent Vandermonde matrix.

Generalizing to higher $n$ and $N_n$ (with $M=\sum_n n N_n<\infty$) the integrand in the definition of $I$ becomes
\be
 \prod_\mathrm{pairs}\left|z_i^{(n)}-z_j^{(m)}\right|^{2nm} = |\det A|^2
\ee
where $A$ is the $M \times M$ confluent Vandermonde matrix
\be
 A_{ij} = \frac{1}{(s-1)!}\left(\frac{\partial}{\partial z_{k}^{(n)}}\right)^{s-1} \left(z_k^{(n)}\right)^{j-1}
\ee
The relation between $n,k,s$ and $i$ is (uniquely) determined by $1\le n$, $1\le k \le N_n$, $1\le s \le n$ and $\ell(n,k)+s=i$ with $\ell(n,k)=\sum_{m=1}^{n-1} m N_m + (k-1)n$.
The result evaluates to
\bea
 &&\prod_\mathrm{pairs}\left|z_i^{(n)}-z_j^{(m)}\right|^{2nm} \nonumber\\
&=& \left|\sum_{\{i\}}\varepsilon_{i_1 \cdots i_M} \prod_n \prod_{k=1}^{N_n} \left[\prod_{s=1}^{n}\frac{1}{(s-1)!}\left(\frac{\partial}{\partial z_{k}^{(n)}}\right)^{s-1} \left(z_{k}^{(n)}\right)^{i_{\ell(n,k)+s}-1}\right]  \right|^2 \\
&=& \left|\sum_{\{i\}}\varepsilon_{i_1 \cdots i_M} \prod_n \prod_{k=1}^{N_n} \left[\prod_{s=1}^{n}\frac{(i_{\ell(n,k)+s}-1)\cdots (i_{\ell(n,k)+s}-s+1)}{(s-1)!} \left(z_{k}^{(n)}\right)^{i_{\ell(n,k)+s}-s}\right]  \right|^2 \nonumber\\
&=& \left|\sum_{\{i\}}\varepsilon_{i_1 \cdots i_M} \prod_n \prod_{k=1}^{N_n} \frac{1}{n!(n-1)!\cdots 1!}\Delta(i_{\ell(n,k)+1},\ldots,i_{\ell(n,k)+n}) \left(z_k^{(n)}\right)^{\sum_{s=1}^n i_{\ell(n,k)+s}} \right|^2 \nonumber
\eea
where $z_k^{(n)}=\exp(it_k^{(n)})$, $|z_k^{(n)}|^2=1$ was used in the last step, and the Vandermonde matrices in the last form are obtained after antisymmetrization. Note that the complicated expression $\ell(n,k)$ is only needed for the pick up the permutation variable $i$ with the correct index.

The constant term is
\bea
 I(N_1,N_2,\ldots) &=&\sum_{\{i\},\{j\}} \varepsilon_{i_1 \cdots i_M} \varepsilon_{j_1 \cdots j_M}\prod_n \prod_{k=1}^{N_n} \frac{1}{\left[n!(n-1)!\cdots 1!\right]^2}\nonumber\\
 &&\times \Delta(i_{\ell(n,k)+1},\ldots,i_{\ell(n,k)+n})\Delta(j_{\ell(n,k)+1},\ldots,j_{\ell(n,k)+n})\nonumber\\
 &&\times  \delta(i_{\ell(n,k)+1}+\cdots+i_{\ell(n,k)+n},j_{\ell(n,k)+1}+\cdots+j_{\ell(n,k)+n})
\eea
where $\delta(i,j)=\delta_{ij}$ is the Kronecker $\delta$-symbol.
For $n=1$ the $\delta$ restrictions give simply $i_k=j_k$. Using these the result ``simplifies'' to
\bea
 \frac{I(N_1,N_2,\ldots)}{N_1!} \!\!&=&\!\! \sum_{S,\{i\},\{j\}} \varepsilon_{i_1 \cdots i_K} \varepsilon_{j_1 \cdots j_K}\prod_{n>1} \prod_{k=1}^{N_n} \frac{1}{\left[n!(n-1)!\cdots 1!\right]^2}\\
 \!\!&&\!\!\times \Delta\left(S(i_{\ell'(n,k)+1}),\ldots,S(i_{\ell'(n,k)+n})\right)\Delta\left(S(j_{\ell'(n,k)+1}),\ldots,S(j_{\ell'(n,k)+n})\right)\nonumber\\
 \!\!&&\!\!\times  \delta\left(\sum_{s=1}^n S(i_{\ell'(n,k)+s}),\sum_{s=1}^n S(j_{\ell'(n,k)+s})\right)\nonumber
\eea
where $K=M-N_1$, the first sum goes over all increasing injections $S:\{1,\ldots,K\} \to \{1,\ldots,M\} $ [so that $i<j \Leftrightarrow S(i)<S(j)$], and $\ell'(n,k) = \ell(n,k)-N_1$.

Due to symmetry, one can add the restrictions $i_{\ell'(n,k)+1}<i_{\ell'(n,k+1)+1}$ (for all $n>1$ and $1\le k<N_n$), and
$i_{\ell'(n,k)+s}<i_{\ell'(n,k)+s+1}$, $j_{\ell'(n,k)+s}<j_{\ell'(n,k)+s+1}$ (for all $n>1$, $k$, and $1\le s < n$) and multiply by the ratio of numbers of terms whence the result becomes
\bea \label{resgen}
 \hat I(N_1,N_2,\ldots)\!\!&=&\!\!\frac{I(N_1,N_2,\ldots)}{\prod_n N_n!}\nonumber\\
 \!\!&=&\!\! \sum_{S,\{i\},\{j\}}\!\!\!' \ \varepsilon_{i_1 \cdots i_K} \varepsilon_{j_1 \cdots j_K}\prod_{n>1} \prod_{k=1}^{N_n} \frac{1}{\left[(n-1)!\cdots 1!\right]^2}\\
 \!\!&&\!\!\times \Delta\left(S(i_{\ell'(n,k)+1}),\ldots,S(i_{\ell'(n,k)+n})\right)\Delta\left(S(j_{\ell'(n,k)+1}),\ldots,S(j_{\ell'(n,k)+n})\right)\nonumber\\
 \!\!&&\!\!\times  \delta\left(\sum_{s=1}^n S(i_{\ell'(n,k)+s}),\sum_{s=1}^n S(j_{\ell'(n,k)+s})\right)\nonumber
\eea
where the prime indicates the presence of the above restrictions. In particular,
\bea \label{resn2}
 \hat I_2(N_1,N_2) &=& \hat I(N_1,N_2,0,0,\ldots) \nonumber\\
 \!\!&=&\!\! \sum_{S,\{i\},\{j\}}\!\!\!' \ \varepsilon_{i_1 \cdots i_{K}} \varepsilon_{j_1 \cdots j_{K}} \prod_{k=1}^{N_2}
  \left(S(i_{2k-1})-S(i_{2k})\right)\left(S(j_{2k-1})-S(j_{2k})\right)\nonumber\\
 \!\!&&\!\!\times  \delta\left(S(i_{2k-1})+S(i_{2k}),S(j_{2k-1})+S(j_{2k})\right)
\eea
where $K=2N_2$ and $\ell'(2,k)=2(k-1)$ was inserted.
We have written computer codes
which evaluate $I$ using the formulae (\ref{resgen}), (\ref{resn2})
for a given (but arbitrary) set of $\{N_n\}$.



\begin{thebibliography}{99}
\bibitem{Witten:1985cc}
  E.~Witten,
  Nucl.\ Phys.\  B {\bf 268}, 253 (1986).

\bibitem{Kiermaier:2007ba}
  M.~Kiermaier, Y.~Okawa, L.~Rastelli and B.~Zwiebach,
  arXiv:hep-th/0701249;
  M.~Schnabl,
  arXiv:hep-th/0701248.

\bibitem{Schnabl:2005gv}
  M.~Schnabl,
  Adv.\ Theor.\ Math.\ Phys.\  {\bf 10}, 433 (2006)
  [arXiv:hep-th/0511286].

\bibitem{susywork}
  T.~Erler,
  arXiv:0704.0930.
  Y.~Okawa,
  arXiv:0704.0936,
  arXiv:0704.3612 [hep-th].

\bibitem{Fuchs:2007yy}
  E.~Fuchs, M.~Kroyter and R.~Potting,
  arXiv:0704.2222 [hep-th].

\bibitem{Coletti:2005zj}
  E.~Coletti, I.~Sigalov and W.~Taylor,
  JHEP {\bf 0508}, 104 (2005)
  [arXiv:hep-th/0505031].

\bibitem{previouswork}
  N.~Moeller and B.~Zwiebach,
  JHEP {\bf 0210}, 034 (2002)
  [arXiv:hep-th/0207107].
  N.~Moeller and M.~Schnabl,
  JHEP {\bf 0401}, 011 (2004)
  [arXiv:hep-th/0304213].
  M.~Fujita and H.~Hata,
  JHEP {\bf 0305}, 043 (2003)
  [arXiv:hep-th/0304163].
  M.~Fujita and H.~Hata,
  Phys.\ Rev.\  D {\bf 70}, 086010 (2004)
  [arXiv:hep-th/0403031].

\bibitem{Kluson:2003xu}
  J.~Kluson,
  JHEP {\bf 0312}, 050 (2003)
  [arXiv:hep-th/0303199].


\bibitem{bsft}
  E.~Witten,
  Phys.\ Rev.\  D {\bf 46}, 5467 (1992)
  [arXiv:hep-th/9208027].
E.~Witten,
  Phys.\ Rev.\  D {\bf 47}, 3405 (1993)
  [arXiv:hep-th/9210065].
 S.~L.~Shatashvili,
  Phys.\ Lett.\  B {\bf 311}, 83 (1993)
  [arXiv:hep-th/9303143].
  S.~L.~Shatashvili,
  Alg.\ Anal.\  {\bf 6}, 215 (1994)
  [arXiv:hep-th/9311177].

\bibitem{Kraus:2000nj}
  P.~Kraus and F.~Larsen,
  Phys.\ Rev.\  D {\bf 63}, 106004 (2001)
  [arXiv:hep-th/0012198].

\bibitem{kmm}
  D.~Kutasov, M.~Marino and G.~W.~Moore,
  JHEP {\bf 0010}, 045 (2000)
  [arXiv:hep-th/0009148].
D.~Kutasov, M.~Marino and G.~W.~Moore,
  arXiv:hep-th/0010108.

\bibitem{Ellwood:2007xr}
  I.~Ellwood,
  arXiv:0705.0013 [hep-th].

\bibitem{SLNT}
  A.~Sen,
  JHEP {\bf 0204}, 048 (2002)
  [arXiv:hep-th/0203211];
  F.~Larsen, A.~Naqvi and S.~Terashima,
  JHEP {\bf 0302} (2003) 039
  [arXiv:hep-th/0212248].

\bibitem{Balasubramanian:2004fz}
  V.~Balasubramanian, E.~Keski-Vakkuri, P.~Kraus and A.~Naqvi,
  Commun.\ Math.\ Phys.\  {\bf 257} (2005) 363
  [arXiv:hep-th/0404039];
  N.~Jokela, E.~Keski-Vakkuri and J.~Majumder,
  Phys.\ Rev.\  D {\bf 73} (2006) 046007
  [arXiv:hep-th/0510205].

\bibitem{Dyson:1962es}
  F.~J.~Dyson,
  J.\ Math.\ Phys.\  {\bf 3} (1962) 140.
  V.~Balasubramanian, N.~Jokela, E.~Keski-Vakkuri and J.~Majumder,
  Phys.\ Rev.\  D {\bf 75} (2007) 063515
  [arXiv:hep-th/0612090].

\bibitem{mehta}
  M.~L.~Mehta, {\em Random Matrices}, 2nd edition, Academic Press
  (1991).

\bibitem{goodwilson}
 K.~G.~Wilson, 
  J.\ Math.\ Phys.\
 {\bf 3} (1962) 1040; I.~J.~Good, 
 J.\ Math.\ Phys.\ {\bf 11} (1970) 1884.

\end{thebibliography}
\end{document}